\documentclass[amsfonts,showpacs,tightenlines,aps,12pt,floatfix,nofootinbib]{revtex4}
\usepackage{bm}
\usepackage{epsfig}
\usepackage{float}
\begin{document}

\thispagestyle{empty}

\title{A Combined NNLO Lattice-Continuum Determination of $L_{10}^r$}

\author{P.A. Boyle}
\affiliation{Physics and Astronomy, University of Edinburgh, Edinburgh
EH9 3JZ, UK}

\author{L. Del Debbio}
\affiliation{Physics and Astronomy, University of Edinburgh, Edinburgh
EH9 3JZ, UK}

\author{N. Garron}
\affiliation{School of Mathematics, Trinity College, Dublin 2, Ireland}

\author{R.J. Hudspith}
\affiliation{Physics and Astronomy, York University, 
Toronto, ON Canada M3J 1P3}

\author{E. Kerrane}
\affiliation{Instituto de F\`isica T\`eorica UAM/CSIC, Universidad
Aut\`onoma de Madrid, Cantoblanco E-28049 Madrid, Spain}

\author{K. Maltman}
\email[]{kmaltman@yorku.ca}
\affiliation{CSSM, Department of Physics, University of Adelaide, Adelaide, 
SA 5005 Australia}
\altaffiliation{Permanent address: Mathematics and Statistics, 
York University, Toronto, ON CANADA M3J 1P3}

\author{J.M. Zanotti}
\affiliation{CSSM, Department of Physics, University of Adelaide, Adelaide
SA 5005 Australia}

\begin{abstract}
The renormalized next-to-leading-order (NLO) chiral low-energy constant, 
$L_{10}^r$, is determined in a complete next-to-next-to-leading-order
(NNLO) analysis, using a combination of lattice and continuum data for the
flavor $ud$ $V-A$ correlator and results from a recent chiral sum-rule 
analysis of the flavor-breaking combination of $ud$ and $us$ 
$V-A$ correlator differences. The analysis also fixes two combinations of
NNLO low-energy constants, the determination of which is crucial to 
the precision achieved for $L_{10}^r$. Using the results of the
flavor-breaking chiral $V-A$ sum rule obtained with current versions
of the strange hadronic $\tau$ branching fractions as input, we find
$L_{10}^r(m_\rho )\, =\, -0.00346(32)$. This result represents the first
NNLO determination of $L_{10}^r$ having all inputs under full
theoretical and/or experimental control, and the best current precision
for this quantity.
\end{abstract}


\maketitle

\section{\label{intro}Introduction}
Chiral perturbation theory (ChPT) provides a framework for implementing, 
in the most general possible way, the constraints placed on the light 
hadronic degrees of freedom by the symmetries and approximate chiral 
symmetry of QCD~\cite{weinberg79,gl84,gl85}. Because the underlying 
arguments are symmetry-based, the resulting effective chiral Lagrangian 
contains as parameters the coefficients (usually called low-energy 
constants, or LECs) multiplying all terms allowed by the symmetry 
constraints. The LECs, which are not determined by the symmetry arguments, 
encode the effects of heavier degrees of freedom such as resonances and are, 
in principle, calculable in the full underlying theory. A key goal in making 
the ChPT framework as predictive as possible is the determination of all LECs 
appearing up to a given order in the chiral expansion. In this paper, we 
focus on the renormalized $SU(3)\times SU(3)$ NLO LEC $L_{10}^r$. 
$L_{10}^r$ is closely related to the $SU(2)\times SU(2)$ LEC $\ell_5^r$, 
and thus also determines the small QCD contribution to 
the $S$-parameter~\cite{spardefn}. 

Previous determinations of $L_{10}^r$, both 
continuum~\cite{dghs98,dominguez,gapp08,bgjmp13}
and lattice~\cite{jlqcdvmanlo,rbcukqcdvmanlo,boylelatt13vma},
were produced by analyses of the low-$Q^2$ behavior of the difference 
of the flavor $ud$ vector ($V$) and axial-vector ($A$) correlators, 
\begin{equation}
\Delta\Pi_{V-A}(Q^2)\equiv \Pi_{ud;V}^{(0+1)}(Q^2)\, -\, 
\Pi_{ud;A}^{(0+1)}(Q^2)\ .
\label{deltapivmadefn}
\end{equation}
Here $\Pi_{ud;V/A}^{(J)}(Q^2)$ are the scalar, spin $J$ components of
the standard flavor $ud$ $V$ and $A$ current-current two-point functions,
$\Pi_{V/A}^{\mu\nu}(Q^2)$, defined by
\begin{eqnarray}
\Pi_{ud;V/A}^{\mu\nu}(q^2)&&\, \equiv\, i\int\, d^4x\, e^{iq\cdot x}
\langle 0\vert T\left( J_{ud;V/A}^\mu (x) J_{ud;V/A}^{\dagger\, \nu}
(0)\right)
\vert 0\rangle
\nonumber\\
&&\,=\, \left( q^\mu q^\nu - q^2 g^{\mu\nu}\right)\, \Pi^{(1)}_{ud;V/A}(Q^2)
\, +\, q^\mu q^\nu\, \Pi_{ud;V/A}^{(0)}(Q^2)\ ,
\label{mink2pt}\end{eqnarray}
where $J_{ud;V}^\mu =V^\mu$ and $J_{ud;A}^\mu =A^\mu$ are the standard 
flavor $ud$ $V$ and $A$ currents, and $Q^2\, =\, -q^2$. The individual 
$\Pi^{(0,1)}_{ud;A}$ have kinematic singularities at $Q^2=0$, but their 
sum, $\Pi^{(0+1)}_{ud;A}$, is kinematic-singularity-free. In what follows, 
the standard notation, $\rho^{(J)}_{ud;V/A}(s)$, with $s\, =\, -Q^2$, will 
be employed for the spectral functions of the $\Pi^{(J)}_{ud;V/A}(Q^2)$. 
$\Delta\rho_{V-A}(s)\equiv \rho^{(0+1)}_{ud;V}(s)-\rho^{(0+1)}_{ud;A}(s)$
is then the spectral function of $\Delta\Pi_{V-A}(Q^2)$.
It is also useful to define the $\pi$-pole-subtracted versions, 
$\overline{\Pi}_{ud;A}$, $\Delta\overline{\Pi}_{V-A}$, 
$\bar{\rho}^{(J)}_{ud;A}$ and $\Delta\bar{\rho}_{V-A}$, 
of $\Pi_{ud;A}$, $\Delta\Pi_{V-A}$, $\rho^{(J)}_{ud;A}$ and
$\Delta\rho_{V-A}$.

As explained in more detail below, the $\rho^{(J)}_{ud;V/A}(s)$
are determinable from experimental hadronic $\tau$ decay distributions. 
Since $\Delta\Pi_{V-A}$ satisfies an unsubtracted dispersion relation,
this allows a continuum determination of $\Delta\Pi_{V-A}(Q^2)$, and 
hence also of $\Delta\overline{\Pi}_{V-A}(Q^2)$, to be achieved.

For $Q^2>0$, $\Delta\Pi_{V-A}(Q^2)$ can also be determined directly on 
the lattice. The results of course depend on the input quark masses used 
in the lattice simulation. The freedom to vary these masses is a useful 
feature for the purpose of determining chiral LECs. We work below with 
lattice ensembles covering a range of $m_u=m_d$ and $m_s$. Ensemble 
$m_\pi$ and $f_\pi$ values then also yield the corresponding 
$\Delta\overline{\Pi}_{V-A}(Q^2)$.

The continuum determination of $\Delta\overline{\Pi}_{V-A}(Q^2)$ is
very precise in the low-$Q^2$ region. Since, to NLO in the chiral 
expansion, the $Q^2$-dependence of $\Delta\overline{\Pi}_{V-A}(Q^2)$ 
is LEC-independent, the continuum results allow a direct determination 
of the only free parameter, $L_{10}^r$, entering the $Q^2$-independent
part of the NLO representation. Unfortunately, there is now clear
evidence that the NLO representation is inadequate in the low-$Q^2$ 
region~\cite{bgjmp13}. The $Q^2$-independent part of the NNLO
representation of $\Delta\overline{\Pi}_{V-A}(Q^2)$, however, involves 
two combinations of NNLO LECs, ${\cal C}_0^r$ and ${\cal C}_1^r$, 
in addition to $L_{10}^r$, making an NNLO determination of $L_{10}^r$ 
impossible without input on the values of these combinations. While the 
coefficients of $L_{10}^r$, ${\cal C}_0^r$ and ${\cal C}_1^r$ depend 
differently on the pseudoscalar masses, the fact that all three 
coefficients are independent of $Q^2$ means the $Q^2$-dependence of 
the continuum $\Delta\overline{\Pi}_{V-A}(Q^2)$ data is of no use in 
disentangling the $L_{10}^r$ contribution. This problem precludes the 
possibility of a fully data-driven continuum NNLO determination of 
$L_{10}^r$.

The fact that the coefficients of $L_{10}^r$, ${\cal C}_0^r$ and 
${\cal C}_1^r$ in the NNLO representation of 
$\Delta\overline{\Pi}_{V-A}(Q^2)$ depend differently on the pseudoscalar 
masses raises the possibility of using lattice data to disentangle
the different $Q^2$-independent contributions, and hence determine
$L_{10}^r$. Unfortunately, because the signal for the lattice two-point 
functions vanish in the limit $Q^2\rightarrow 0$, errors on the lattice 
data for $\Delta\overline{\Pi}_{V-A}(Q^2)$ are large in the low-$Q^2$ 
region, too large, as it turns out, to allow a purely lattice NNLO
analysis to be carried out. 

In this paper, we show how the complementary advantages of the 
continuum and lattice approaches can be combined to produce an NNLO 
determination of $L_{10}^r$ which would not be possible using either 
approach alone. The rest of the paper is organized as follows. In 
Sec.~\ref{sec2}, we expand on the background outlined above, providing 
technical details and notation of relevance to the analysis to follow. 
In Sec.~\ref{sec3}, we recall briefly certain key results from the 
continuum analysis of $\Delta\overline{\Pi}_{V-A}(Q^2)$ reported in 
Ref.~\cite{bgjmp13}, also of relevance to the analysis below. Details 
of the lattice simulations are provided in Sec.~\ref{sec4a}, and an 
outline of the procedure for generating the $V$ and $A$ two-point 
functions on the lattice in Sec.~\ref{sec4b}. Sec.~\ref{sec4c}
presents the resulting $\Delta\overline{\Pi}_{V-A}(Q^2)$ lattice data,
and provides further detail on the problems encountered in attempting 
to carry out a NNLO analysis of the lattice data alone. In 
Sec.~\ref{sec5}, we discuss how to combine lattice data, continuum 
data, and a continuum constraint on $\Delta\overline{\Pi}_{V-A}(0)$
to produce determinations of all three LECs $L_{10}^r$, ${\cal C}_0^r$ 
and ${\cal C}_1^r$, and how to further improve these determinations by 
incorporating a constraint from the recent inverse-moment finite
energy sum rule analysis of the flavor-breaking difference
of $ud$ and $us$ V-A correlators reported in Ref.~\cite{kmimfesr13}. 
Finally, in Sec.~\ref{sec6}, we provide a brief summary, and discussion 
of our results.

\section{\label{sec2}Background}
Continuum results for $\Delta\Pi_{V-A}(Q^2)$ can be obtained via the
unsubtracted dispersion relation
\begin{equation}
\Delta\Pi_{V-A}(Q^2)\, =\, \int_0^\infty ds\, {\frac{\Delta\rho_{V-A}(s)}
{s+Q^2}}\ .
\label{dispvma}\end{equation}
The corresponding result for $\Delta\overline{\Pi}_{V-A}(Q^2)$ is 
obtained by replacing $\Delta\Pi_{V-A}$ with 
$\Delta\overline{\Pi}_{V-A}(Q^2)$, $\Delta\rho_{V-A}$ with
$\Delta\bar{\rho}_{V-A}$ and the lower limit on the RHS 
with the continuum threshold, $4m_\pi^2$, in Eq.~(\ref{dispvma}).

For $s<m_\tau^2$, the $\rho^{(J)}_{ud;V/A}(s)$ are accessible 
experimentally through the normalized differential non-strange 
hadronic $\tau$ decay distributions, $dR_{ud;V/A}/ds$, where 
\begin{eqnarray}
R_{ud;V/A}\, &&\equiv\, \Gamma [\tau^- \rightarrow \nu_\tau
\, {\rm hadrons}_{ud;V/A}\, (\gamma )]/ \Gamma [\tau^- \rightarrow
\nu_\tau e^- {\bar \nu}_e (\gamma)]\ .
\end{eqnarray}
Explicitly~\cite{tsai}
\begin{eqnarray}
&&{\frac{dR_{ud;V/A}}{ds}}\, =\, {\frac{12\pi^2\vert V_{ud}\vert^2 S_{EW}}
{m_\tau^2}}\, \left[ w_\tau \left( y_\tau \right)\, \rho_{ud;V/A}^{(0+1)}(s)\,
-\, w_L \left( y_\tau \right)\, \rho_{ud;V/A}^{(0)}(s) \right]
\label{basictaudecay}\end{eqnarray}
with $y_\tau = s/m_\tau^2$, $w_\tau (y)=(1-y)^2(1+2y)$, $w_{L}(y)=2y(1-y)^2$, 
$S_{EW}$ a known short-distance electroweak correction~\cite{erler},
and $V_{ud}$ the flavor $ud$ element of the CKM matrix. 

Apart from the $\pi$ pole contribution to $\rho_{ud;A}^{(0)}(s)$, which is 
not chirally suppressed, all other contributions to $\rho_{ud;V/A}^{(0)}(s)$ 
are proportional to $(m_d\mp m_u)^2$, and hence numerically negligible. 
The combination $\rho^{(0+1)}_{ud;V+A}(s)$ is thus directly determinable 
from the non-strange differential decay distribution. To form the $V-A$
difference requires a $V/A$ separation. The bulk of this separation can be 
performed using G-parity, which is unambiguous for $n\, \pi$ states. The 
main remaining uncertainty, in the region covered by the $\tau$ decay data, 
is that associated with contributions to the inclusive spectrum from 
$K\bar{K}\pi$ states, for which G-parity cannot be used. The separation in 
this case could, in principle, be accomplished through a relatively simple 
angular analysis~\cite{km92}, but this has yet to be done. The publicly 
accessible OPAL~\cite{opalud99} versions of the inclusive $V$ and $A$ spectral 
distributions have been obtained assuming a maximally conservative, fully 
anticorrelated $50\pm 50\%$ $V/A$ breakdown of the $\bar{K}K\pi$ and much 
smaller $\bar{K}K\pi\pi$ contributions. ALEPH data is also available, 
the 2005 version employing an improved $V/A$ separation of 
$\bar{K}K\pi\pi$ contributions obtained using CVC and isovector 
$\bar{K}K\pi$ electroproduction cross-section results~\cite{alephud}.

The continuum results we employ below are those reported in 
Ref.~\cite{bgjmp13}, obtained using the updated version of the OPAL 
data~\cite{opalud99} detailed in Ref.~\cite{dv72}. (An error in the 
publicly accessible version of the ALEPH covariance matrix prevented the 
use of the nominally higher-precision ALEPH data~\cite{dv7tau10},
the recently released corrected version~\cite{correctedaleph13} 
having not been posted until after the work reported here was completed.) 
The $\tau$ decay data covers the region only up to $s=m_\tau^2$
in the dispersive representation. Above this point, $\Delta\rho_{V-A}(s)$ 
was obtained using a phenomenologically successful, experimentally 
constrained model for duality violations (DVs) investigated extensively in 
Ref.~\cite{dv72,dv71}. In the region of low $Q^2$ relevant to the chiral 
analysis, the resulting DV contributions to the dispersive 
result for $\Delta\overline{\Pi}_{V-A}(Q^2)$ are numerically very small, 
making the result an essentially entirely experimentally determined one.
The key output from this analysis, for our purposes below, is the very 
precise determination~\cite{bgjmp13},
\begin{equation}
\Delta\overline{\Pi}_{V-A}(0)\, =\, 0.0516(7)\ .
\label{pivmaudzero}\end{equation}

The chiral expansion of $\Delta\overline{\Pi}_{V-A}(Q^2)$ to NLO
has the form~\cite{gl84,abt00}
\begin{equation}
\left[ \Delta\overline{\Pi}_{V-A}(Q^2)\right]_{NLO}\, =\, 
-8\, L_{10}^r \, +\, {\cal R}_{NLO}(Q^2)\ ,
\label{nlopibarvma}\end{equation}
where $Q^2\, =\, -q^2$, and ${\cal R}_{NLO}(Q^2)$, which contains all 
contributions from 1-loop graphs with only leading-order (LO) vertices, 
is completely fixed, for a given $Q^2$, by the $\pi$ and $K$ masses and 
the chiral renormalization scale $\mu$.
$L_{10}^r$ of course also depends on $\mu$. At NLO, 
$\Delta\overline{\Pi}_{V-A}(0)$ is thus determined by the single 
parameter $L_{10}^r$, and, as noted above, 
a determination of $\Delta\overline{\Pi}_{V-A}(0)$ 
translates into an NLO determination of $L_{10}^r$. 

\begin{figure}[H]
\caption{\label{fesrcontour}The contour underlying the chiral sum rules of
Eq.~(\ref{imsrbasic})}
\centering
{\rotatebox{270}{\mbox{
\includegraphics[width=2in]{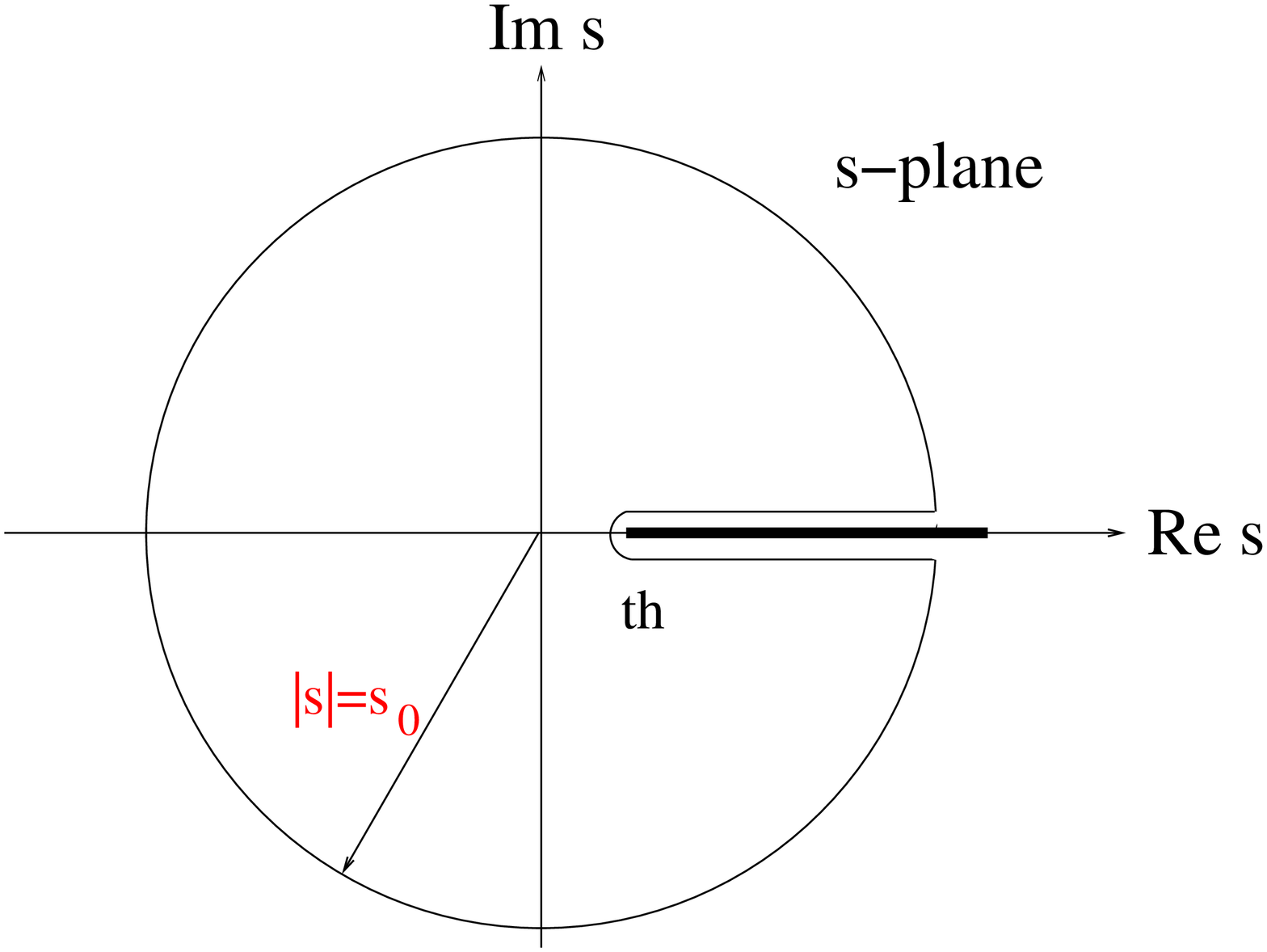}
}}}
\end{figure}

$\Delta\overline{\Pi}_{V-A}(0)$ can be obtained either from the
dispersive representation, or through the use of inverse-moment finite
energy sum rules (IMFESRs). These are sum rules based on the integration,
over the contour shown in Fig.~\ref{fesrcontour}, of the product 
$w(s)\, \tilde{\Pi} (s)$, where $w(s)$ is any function analytic in the 
region of the contour and $\tilde{\Pi} (s)\equiv\Pi (Q^2)$ (with $Q^2\, =\, 
-s$) any correlator free of kinematic singularities.
With $\rho (s)$ the spectral function of $\Pi (Q^2)$, the resulting IMFESR
relation is
\begin{eqnarray}
&&w(0)\, \Pi (0) \, =\, {\frac{1}{2\pi i}}\,\oint_{\vert s\vert = s_0} ds\,
{\frac{w(s)}{s}}\, \Pi (Q^2)\ +\ \int_{th}^{s_0}ds\, {\frac{w(s)}{s}}\,\rho
(s)\
\ ,
\label{imsrbasic}\end{eqnarray}
where $th$ is the threshold shown in Fig.~\ref{fesrcontour}.
For large enough $s_0$, the Operator Product Expansion (OPE) 
representation of $\Pi (s)$ can be used
in evaluating the first term on the RHS. The IMFESR relation is based 
on the same analyticity properties as the basic dispersion relation, 
the information on the integral from $s_0$ to $\infty$ in the dispersive
representation being replaced, in the IMFESR approach, by the OPE 
approximation to the integral around the circle $\vert s\vert =s_0$. 
The added advantage of the IMFESR formulation lies in the freedom to 
choose the weight $w(s)$ in such a way as to improve various features 
of the evaluation of the RHS of Eq.~(\ref{imsrbasic}). 

Early continuum NLO determinations of $L_{10}^r$, using the IMFESR 
approach, were performed in Refs.~\cite{dghs98,dominguez}. Two NLO 
lattice determinations, based on analyses of low-Euclidean-$Q^2$ lattice 
data for $\Delta\Pi_{V-A}(Q^2)$, also exist~\cite{jlqcdvmanlo,rbcukqcdvmanlo}.
The only $Q^2$-dependence of $\Delta\overline{\Pi}_{V-A}(Q^2)$ at NLO lies 
in the loop contribution, ${\cal R}_{NLO}(Q^2)$. It is now known that this 
dependence provides a very poor representation of the actual low-$Q^2$ 
behavior of $\Delta\overline{\Pi}_{V-A}(Q^2)$~\cite{bgjmp13} (a similar 
observation was also made regarding the NLO representation of the $ud$ 
V correlator, $\Pi_{ud;V}(Q^2)$, relevant to lattice determinations of 
the LO hadronic vacuum polarization contribution to the muon anomalous
magnetic moment~\cite{ab06}). This raises obvious questions for the 
earlier NLO $L_{10}^r$ determinations.

The NNLO representation of $\Delta\overline{\Pi}_{V-A}(Q^2)$, needed to 
extend the NLO continuum dispersive/IMFESR determinations to NNLO,
has the form~\cite{abt00}
\begin{equation}
\left[ \Delta\overline{\Pi}_{V-A}(Q^2)\right]_{NNLO} \, =\,
{\cal R}_{NNLO}(Q^2) \, +\, c_{9}(Q^2) L^r_9\, +\,
c_{10}L^r_{10} \, +\, {\cal C}^r_0\, +\, {\cal C}^r_1\, -\, 16 C_{87}^rQ^2\ , 
\label{udvmannlo}\end{equation}
where ${\cal R}_{NNLO}(Q^2)$ is the sum of 1- and 2-loop contributions
involving only LO vertices,
\begin{eqnarray}
&&c_{10}\, =\, -8\left(1-8 \mu_\pi -4\mu_K\right)\ ,
\label{l9l10udvmacoeffs}\end{eqnarray}
with $\mu_P = {\frac{m_P^2}{32\pi^2f_\pi^2}}\, \log\left({\frac{m_P^2}
{\hat{\mu}^2}}\right)$ the usual chiral logarithm and
$f_\pi\simeq 92.2\ MeV$, $c_9(Q^2)$ involves both chiral log and 
standard 1-loop, 2-propagator contributions, and
\begin{eqnarray}
{\cal C}^r_0\, &&=\, 32m_\pi^2\left[
C_{12}^r-C_{61}^r+C_{80}^r\right]\nonumber\\
{\cal C}^r_1\, &&=\, 32\left( m_\pi^2+2m_K^2\right)\,
\left[ C_{13}^r-C_{62}^r+C_{81}^r\right]\ .
\label{nnlolecudvma}\end{eqnarray}
The $C_k^r$ here are the renormalized, dimensionful NNLO LECs defined in 
Ref.~\cite{bce99}. The expression for ${\cal R}_{NNLO}(Q^2)$, which is 
rather lengthy and hence not presented here, is readily reconstructed 
from the results quoted in Sections 4, 6 and Appendix B of Ref.~\cite{abt00},
as is that for $c_9(Q^2)$. $c_{10}$ and, for given $Q^2$, 
${\cal R}_{NNLO}(Q^2)$ and $c_9(Q^2)$ are 
all fixed by the chiral scale $\mu$ and pseudoscalar decay constants and 
masses. The NNLO LECs in ${\cal C}^r_0$ are LO in $1/N_c$, while those in
${\cal C}^r_1$ are $1/N_c$-suppressed. 

The NLO LEC $L_9^r$ has been accurately determined in an NNLO analysis 
of $\pi$ and $K$ electromagnetic form factors~\cite{bt02}, and will be 
considered known in what follows. To simplify notation, we combine
the known terms on the RHS of (\ref{udvmannlo}), defining
\begin{equation}
\hat{\cal R}_{NNLO}(Q^2)\equiv {\cal R}_{NNLO}(Q^2)\, +\, c_9(Q^2)L_9^r\ .
\label{Rhatdefn}\end{equation}
Even with $L_9^r$ known, the NNLO representation of 
$\Delta\overline{\Pi}_{V-A}(0)$ depends on the two NNLO LEC combinations, 
${\cal C}_0^r$ and ${\cal C}_1^r$, in addition to $L_{10}^r$. 
$L_{10}^r$ is thus no longer fixed by a determination
of $\Delta\overline{\Pi}(0)_{V-A}$. Considering the $Q^2$-dependence of 
$\Delta\overline{\Pi}_{V-A}(Q^2)$ does not help resolve this problem
since the terms involving $L_{10}^r$, ${\cal C}^r_0$ and ${\cal C}^r_1$ 
in Eq.~(\ref{udvmannlo}) are all $Q^2$-independent. Additional input on 
${\cal C}_0^r$ and ${\cal C}_1^r$ is thus required to achieve a 
determination of $L_{10}^r$. 

The ${\cal C}_0^r$ contribution to $\Delta\overline{\Pi}_{V-A}(0)$ is 
proportional to $m_\pi^2$ and expected to be small. In 
Ref.~\cite{gapp08}, existing determinations of $C_{12}^r$~\cite{jopc12} 
and $C_{61}^r$~\cite{dk00}, and resonance ChPT (RChPT) estimates for 
$C_{80}^r$~\cite{abt00,up08}, were used to confirm this expectation.
Neglect of the ${\cal C}_1^r$ contribution is far less safe since
the ratio, $(m_\pi^2+2m_K^2)/m_\pi^2\simeq 26$, of the prefactors in
${\cal C}_1^r$ and ${\cal C}_0^r$ more than compensates for the 
$1/N_c$ suppression of the NNLO LECs $C_{13,62,81}^r$ appearing
in ${\cal C}_1^r$. Even more problematic is the fact that previous 
determinations exist for none of $C_{13,62,81}^r$, and that standard 
RChPT approaches yield no estimates for any of these LECs. This problem 
was dealt with in Ref.~\cite{gapp08} by assigning to the $1/N_c$-suppressed 
combination $C_{13}^r(\mu_0)-C_{62}^r(\mu_0)+C_{81}^r(\mu_0)$ 
(with $\mu_0$ the conventional chiral scale choice $\mu =0.77\ GeV$) 
a central value zero and error equal to $1/3=1/N_c$ of the value of 
the corresponding non-$1/N_c$-suppressed combination 
$C_{12}^r(\mu_0)-C_{61}^r(\mu_0)+C_{80}^r(\mu_0)$ appearing in
${\cal C}_0^r$. Given the rather strong cancellations in the latter
combination, this assumption is a far from conservative one. 
The uncertainty on the result for $L_{10}^r$ obtained after implementing
this assumption in the NNLO representation of 
$\Delta\overline{\Pi}_{V-A}(0)$ turns
out to be completely dominated by the assumed error on
$C_{13}^r(\mu_0)-C_{62}^r(\mu_0)+C_{81}^r(\mu_0)$. Improvements to
this unsatisfactory situation can be achieved only through 
an independent determination of ${\cal C}_1^r$.

The fact that the coefficients of $L_{10}^r$,
$C_{12}^r(\mu_0)-C_{61}^r(\mu_0)+C_{80}^r(\mu_0)$ and
$C_{13}^r(\mu_0)-C_{62}^r(\mu_0)+C_{81}^r(\mu_0)$ in 
Eq.~(\ref{udvmannlo}) depend differently on the pseudoscalar meson 
masses suggests disentangling the $L_{10}^r$, ${\cal C}_0^r$ and 
${\cal C}_1^r$ contributions to $\Delta\overline{\Pi}_{V-A}(Q^2)$ 
might be possible on the lattice, where variations in the pseudoscalar 
masses are easily accomplished through variations in the input quark 
masses. This paper shows how this possibility can be realized
practically in an analysis using a combination of lattice and continuum 
results.

\section{\label{sec3}Information from the Continuum Analysis of
$\Delta\overline{\Pi}_{V-A}(Q^2)$}
The LECs $L_{10}^r$, ${\cal C}_0^r$ and ${\cal C}_1^r$ are 
very tightly constrained by (\ref{pivmaudzero}). 
Inputting the results of Ref.~\cite{abt00} for ${\cal R}_{NNLO}(0)$,
and $L_9^r(\mu_0)=0.00593(43)$ from Ref.~\cite{bt02}, 
this constraint takes the form~\cite{bgjmp13}
\begin{equation}
L_{10}^r(\mu_0 )\, -\,
0.0822\left[{\cal C}^r_0(\mu_0)+{\cal C}^r_1(\mu_0)\right]\, =\, 
-0.004098(59)_{exp}(74)_{L_9^r}
\label{nnlopibarvmaq2eq0}\end{equation}
where the subscripts $exp$ and $L_9^r$ label contributions to the
error on the RHS associated with that in
(\ref{pivmaudzero}), and the uncertainty on $L_9^r(\mu_0)$,
respectively. 

Other information from the continuum analysis of Ref.~\cite{bgjmp13}
relevant to the analysis below concerns the range of validity of the NNLO
representation. Crucial to the use of the lattice data is the ability 
to perform a chiral fit to the lattice data at non-zero Euclidean $Q^2$ 
and then use the results of that fit to reliably extrapolate to $Q^2=0$.
This needs to be done for a range of pseudoscalar meson masses in
order to allow the contributions of $L_{10}^r$, ${\cal C}_0^r$ and
${\cal C}_1^r$ to $\Delta\overline{\Pi}_{V-A}(0)$ to be disentangled.
One thus needs to restrict one's attention to lattice data at $Q^2$ for
which the chiral representation being employed is reliable, and, of particular
importance for our purposes, for which one knows the fit will
produce a reliable determination of the $Q^2$-independent part of
the representation, or, equivalently, $\Delta\overline{\Pi}_{V-A}(0)$.

As we will see in the next section, lattice errors on 
$\Delta\overline{\Pi}_{V-A}(Q^2)$ turn out to be too large to allow
the range of validity to be assessed using lattice data alone. 
Moreover, because, for Euclidean $Q^2$, $Q^2=0$ requires all components
of $Q$ to be zero, the signal for the current-current two-point
function vanishes on the lattice as $Q^2\rightarrow 0$. This means
that $\Delta\overline{\Pi}_{V-A}(0)$ cannot be measured directly on
the lattice, and that errors on $\Delta\overline{\Pi}_{V-A}(Q^2)$
are necessarily large for very low $Q^2$. 

The continuum dispersive approach, which produces significantly smaller errors 
on $\Delta\overline{\Pi}_{V-A}(Q^2)$ in the low-$Q^2$ region relevant 
to the chiral analysis, and has no problem in determining 
$\Delta\overline{\Pi}_{V-A}(0)$ directly, is complementary in this regard.
From Eq.~(\ref{udvmannlo}), it is evident that, since
${\cal R}_{NNLO}(Q^2)$ and $c_9(Q^2)L_9^r$ are known, the NNLO
form is characterized by two parameters, $C_{87}^r$ and the 
combination $c_{10}L_{10}^r+{\cal C}_0^r+{\cal C}_1^r$.
In Ref.~\cite{bgjmp13} it was found that the NNLO form produces
a very accurate fit to the continuum data in a fit window covering
the range from $Q^2=0$ to $\sim 0.1\ GeV^2$, one which, moreover,
nicely reproduces the known value of $\Delta\overline{\Pi}_{V-A}(0)$.
Extending the upper edge of the fit window beyond $\sim 0.1 \ GeV^2$,
one starts to see signs of curvature with respect to $Q^2$ beyond that
present in the NNLO representation. This is especially evident in a drift 
in the fitted value for $C_{87}^r(\mu_0)$ as the fit window is opened up,
but also shows up in an accompanying small downward drift in the fitted 
result for $\Delta\overline{\Pi}_{V-A}(0)$~\cite{bgjmp13}. Curvature of 
$\Delta\overline{\Pi}_{V-A}(Q^2)$ with respect to $Q^2$, beyond that 
produced by the nearly linear ${\cal R}_{NNLO}(Q^2)$ contribution, would 
first appear at NNNLO in the chiral expansion, where it would be 
represented by a term of the form $CQ^4$, with the coefficient
$C$ independent of the pseudoscalar meson masses at this order. 
Adding such a term to the NNLO form, stabilizes the fit results
for $C_{87}^r$ as a function of the upper edge of the fit window, and 
restores the success of the resulting representation in
reproducing the known value of $\Delta\overline{\Pi}_{V-A}(0)$
for fit windows with upper edges extending up to 
$\sim 0.3\ GeV^2$~\cite{bgjmp13}.
This information motivates the restriction on the lattice data 
to be used in our analysis, described in the next section, to 
$Q^2<0.3 \ GeV^2$.

\section{\label{sec4}The Lattice Data for $\Delta\overline{\Pi}_{V-A}(Q^2)$}
\subsection{\label{sec4a}Simulation Details}
We consider data on $\Delta\overline{\Pi}_{V-A}(Q^2)$ obtained from five 
RBC/UKQCD $n_f=2+1$ domain wall fermion (DWF) ensembles, three with 
Iwasaki gauge action, inverse lattice spacing $1/a\, =\, 2.31\ GeV$, 
pion masses $m_\pi\, =\, 293$, $349$ and $399\ MeV$, and 
$m_\pi L=4.1$, $4.8$, $5.5$, respectively, and two with 
Iwasaki+DSDR gauge action, $1/a\, =\, 1.37\ GeV$, $m_\pi\, =\, 171$ 
and $248\ MeV$ and $m_\pi L=4.0,\ 5.5$, respectively. 

\begin{table}[H]
\centering
\begin{tabular}{c|c|c|c|c|c|c|c|c|c}
  $Ensemble$ & $V$ & $\beta$ & $a^{-1}\, {[GeV]}$ &
  $Q^2_{min}\, {[GeV^2]}$ & $am_s$ &
  $am_u$ & $m_\pi\, {[GeV]}$ & $m_K\, {[GeV]}$ &
  $F_\pi\, {[GeV]}$ \\\hline\hline
  E1 & $32^3\times64$ & 1.75 & 1.37(1) & 0.018 & 0.045 & 0.001  & 0.171(1) &
  0.492(1) & 0.130(2)\\
  E2 & $32^3\times64$ & 1.75 & 1.37(1) & 0.018 & 0.045 & 0.0042 & 0.248(1) &
  0.509(1) & 0.139(2)\\
  E3 & $32^3\times64$ & 2.25 & 2.31(4) & 0.05 & 0.03 & 0.004    & 0.293(1) &
  0.561(1) & 0.142(1)\\
  E4 & $32^3\times64$ & 2.25 & 2.31(4) & 0.05 & 0.03 & 0.006    & 0.349(1) &
  0.578(1) & 0.148(1)\\
  E5 & $32^3\times64$ & 2.25 & 2.31(4) & 0.05 & 0.03 & 0.008    & 0.399(1) &
  0.596(1) & 0.154(1)
\end{tabular}
\caption{Parameters of the lattice ensembles used in our
  study. $m_\pi$, $m_K$ and $F_\pi$ are from \cite{ainv231}
  ($E3$-$E5$) and \cite{ainv137} ($E1,\,E2$).}
\label{tab:param}
\end{table}

The simulation parameters for the lattice calculations are summarized in 
Table~\ref{tab:param}. Along with the bare lattice simulation parameters, 
we also list the associated values of $m_\pi$, $m_K$ and $F_\pi \equiv 
\sqrt{2}f_\pi$, as well as the minimum $Q^2$ value attainable for each 
lattice, which is governed by its physical volume. Further details of 
the simulations for the three fine and two coarse ensembles may be found 
in Refs.~\cite{ainv231} and ~\cite{ainv137}, respectively.

The fine ensembles provide only three $Q^2$ values in the region 
$Q^2<0.3\ GeV^2$ employed in the current analysis. At the lowest of these,
$Q^2\sim 0.05\ GeV^2$, the errors on $\Delta\overline{\Pi}_{V-A}(Q^2)$, 
moreover, are so large that the result at this $Q^2$ plays no functional role 
in the analysis. The constraints obtained using these ensembles thus
come from the two intermediate-$Q^2$ points. The coarse ensembles have 
improved low-$Q^2$ coverage, providing seven $Q^2$ values below
$0.3\ GeV^2$, four with errors small enough that the corresponding
data plays a role in the analysis. 

\subsection{\label{sec4b}The Current-Current Two-Point Functions
on the Lattice}
In this work we will need to consider the standard lattice
current-current two-point correlation functions,
defined, in momentum space, for the $V$ and $A$ currents, by
\begin{eqnarray}
  \Pi^{\mu\nu}_{ud;V}(Q^2)&&\equiv Z_V \sum_{x} e^{\mathrm{i} Q\cdot x}
  \langle 0|\mathcal{V}^\mu(x)V^\nu(0)|0\rangle, 
\label{eq:Vpidef}\\
  \Pi^{\mu\nu}_{ud;A}(Q^2)&&\equiv Z_A \sum_{x} e^{\mathrm{i} Q\cdot x}
  \langle 0|\mathcal{A}^\mu(x)A^\nu(0)|0\rangle,
\label{eq:Apidef}
\end{eqnarray}
where we use the standard flavor $ud$ DWF conserved vector 
($\mathcal{V}^\mu$) and axial-vector ($\mathcal{A}^\mu$) 
currents~\cite{Furman:1994ky} at the sink. At the source we use 
the corresponding local currents, $V^\mu$ and $A^\mu$, and have 
hence included the vector and axial-vector renormalization constants, 
$Z_V$ and $Z_A$, in Eqs.~(\ref{eq:Vpidef}) and (\ref{eq:Apidef}). 
The values of $Z_V$ and $Z_A$ for each of our ensembles were 
determined in \cite{ainv231,ainv137}.

The two-point functions in Eqs.~(\ref{eq:Vpidef}) and
(\ref{eq:Apidef}) can be decomposed into longitudinal ($J=0$) and
transverse ($J=1$) components,
\begin{equation}
  \label{eq:pi1}
  \Pi^{\mu\nu}_{ud;V/A}=\left(Q^2\delta_{\mu\nu}-Q_\mu Q_\nu\right)
  \Pi^{(1)}_{ud;V/A}(Q^2) - Q_\mu Q_\nu\Pi^{(0)}_{ud;V/A}(Q^2)\,.
\end{equation}
On the lattice momenta are discretised, $Q_\mu=\frac{2\pi
  n_\mu}{L_\mu}$ where $n_\mu$ is a 4-tuple of integers, and $L_\mu$
is the length of the lattice in the $\mu$ direction. In what follows, we
will use the lattice momentum
\begin{equation}
 \hat{Q}_\mu=\frac{2}{a}\sin\left(\frac{\pi n_\mu}{L_\mu}\right).
\end{equation}
and associate the quantity $\hat{Q}^2=\sum_\mu\hat{Q}_\mu^2$ with the
continuum spacelike squared-momentum $Q^2$.

The two-point correlators used here are the same as those used 
previously in studies of the QCD S-parameter~\cite{rbcukqcdvmanlo}
and the hadronic contribution to the anomalous magnetic moment of the 
muon~\cite{Boyle:2011hu}, and  we refer the interested reader to those papers
for more technical details.

\subsection{\label{sec4c}The Lattice V-A Results}
In Table~\ref{tab:param} we provide the values of $m_\pi$, $m_K$
and $F_\pi$ for each of the lattice ensembles. These are needed
both for the $\pi$-pole subtraction, required to convert
$\Delta\Pi_{V-A}(Q^2)$ to $\Delta\overline{\Pi}_{V-A}(Q^2)$, and
in evaluating the 1- and 2-loop contributions to the
NNLO representation of $\Delta\overline{\Pi}_{V-A}(Q^2)$ for
each of the ensembles. The error on the $\pi$-pole subtraction,
produced by uncertainties in the ensemble values of $F_\pi$ and
$m_\pi$, and that on $\Delta\Pi_{V-A}(Q^2)$, are treated as
independent in computing the error on $\Delta\overline{\Pi}_{V-A}(Q^2)$.
Results for further observables for the 
three fine ensembles may be found in Ref.~\cite{ainv231} and  
for the two coarse ensembles in Ref.~\cite{ainv137}. In what
follows, we identify individual ensembles using the 
labels ($E1$--$E5$) introduced to specify them in the Table.

\begin{figure}[!htb]
  \begin{minipage}[t]{0.46\linewidth}
{\rotatebox{270}{\mbox{
\includegraphics[width=0.9\textwidth]
{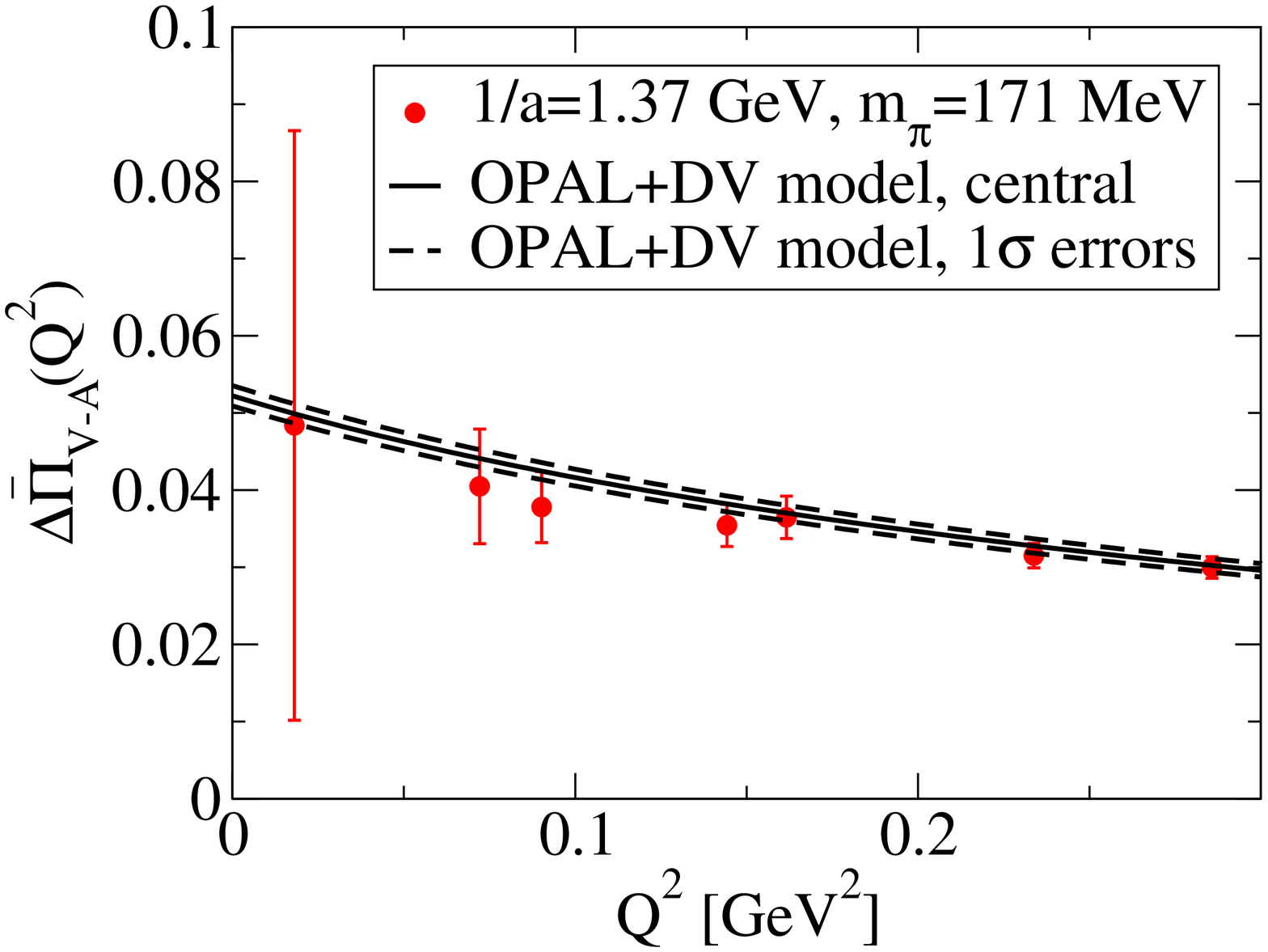}
}}}
  \end{minipage}
\hfill
  \begin{minipage}[t]{0.46\linewidth}
{\rotatebox{270}{\mbox{
\includegraphics[width=0.9\textwidth]
{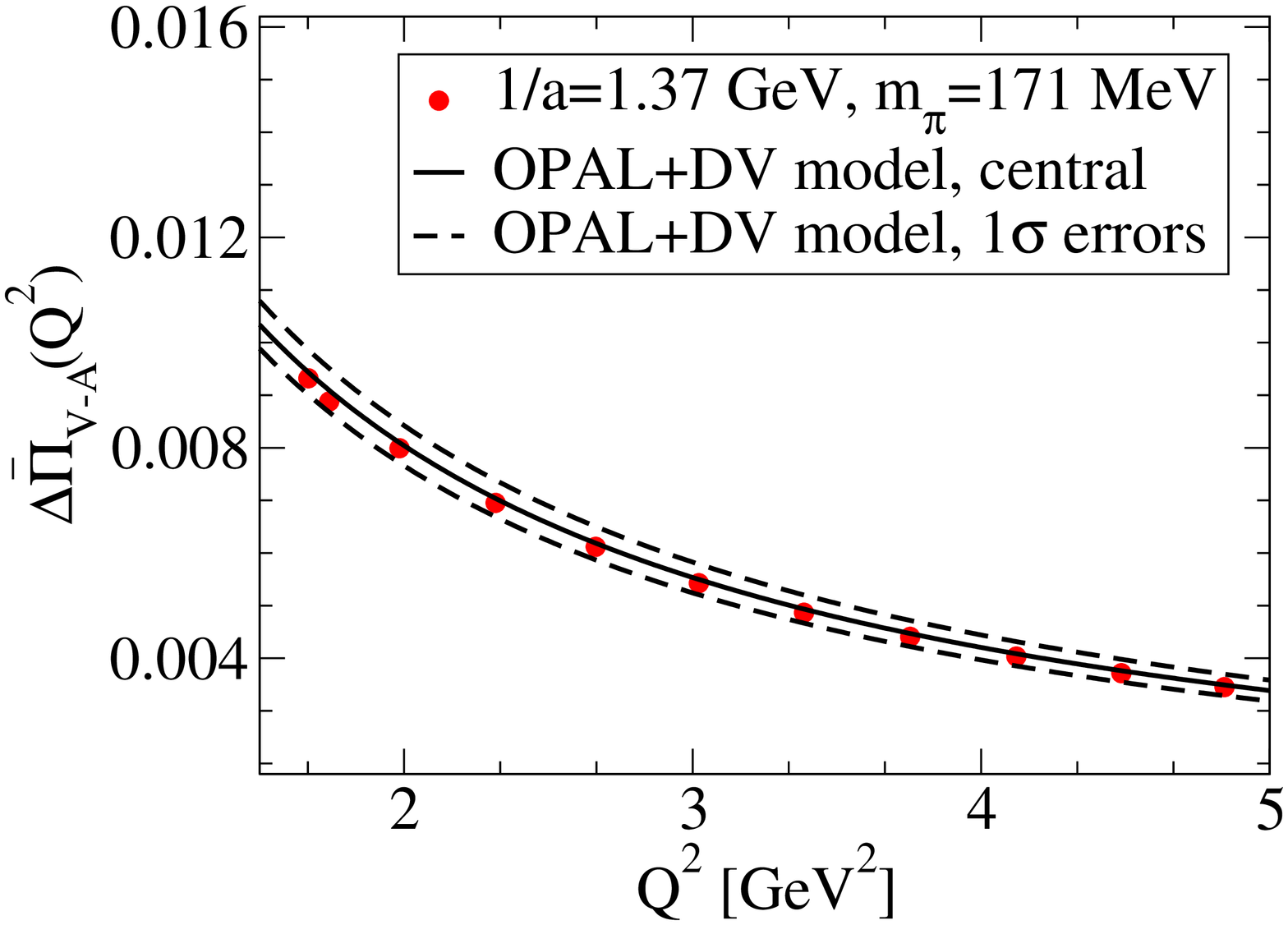}
}}}
  \end{minipage}
  \caption{  \label{continlattcomparisons}Comparison of continuum and 
$1/a=1.37\ GeV$, $m_\pi =171\ MeV$ ensemble lattice results for 
$\Delta\overline{\Pi}_{V-A}(Q^2)$ in the low-$Q^2$ (left panel)
and high-$Q^2$ (right panel) regions}
\end{figure}

A comparison of the continuum (dispersive) results for 
$\Delta\overline{\Pi}_{V-A}(Q^2)$ to those for ensemble $E1$ (whose 
$m_\pi$ value, $171 \ MeV$, lies closest to the physical one) are 
shown in Fig.~\ref{continlattcomparisons}. We would expect these to be
in good agreement, since the $\pi$ pole contribution, which
depends more sensitively on $m_\pi$, has been subtracted in forming
$\Delta\overline{\Pi}_{V-A}(Q^2)$. The left panel shows the
comparison in the low-$Q^2$ chiral fit region, $0<Q^2<0.3\ GeV^2$,
the right panel the comparison for $Q^2\sim$ a few $GeV^2$.
The agreement in both regions is good, suggesting lattice artifacts
are well under control.

\begin{figure}[!htb]
{\centering
{\rotatebox{270}{\mbox{
\includegraphics[width=4in]
{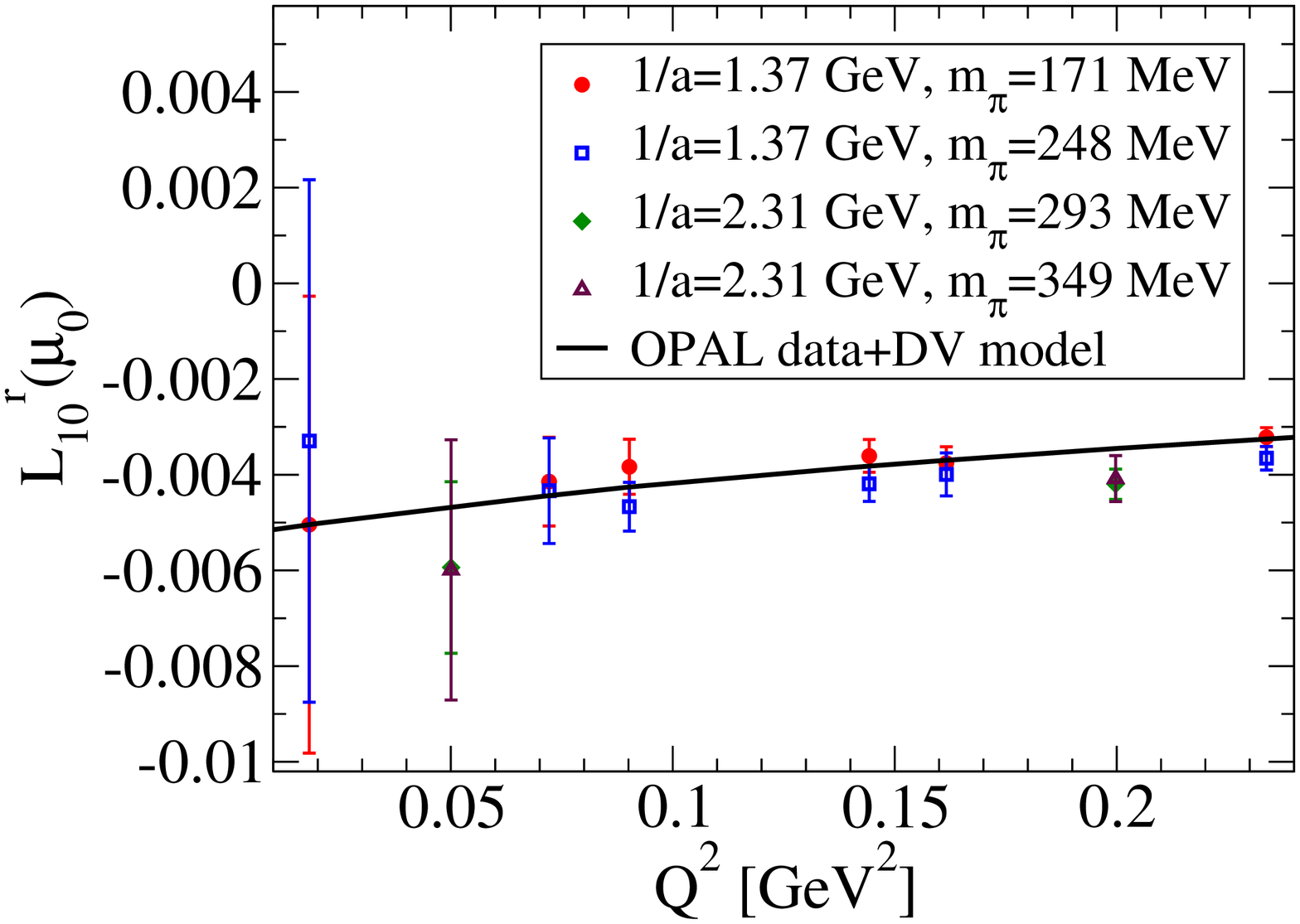}
}}}
}
\caption{\label{nlol10}Point-by-point determinations of 
$L_{10}^r(\mu_0)$, with $\mu_0 =0.77\ GeV$, obtained 
assuming the validity of the NLO form, Eq.~(\ref{nlopibarvma}), for 
$\Delta\overline{\Pi}_{V-A}(Q^2)$. Points with error bars are obtained
from the lattice data discussed in the text, while the continuous
curve results from applying the NLO form to the continuum dispersive
results for $\Delta\overline{\Pi}_{V-A}(Q^2)$.}
\end{figure}

Fig.~\ref{nlol10} illustrates the problems that would be encountered
if one attempted an NNLO analysis involving lattice data alone. The 
figure shows the values of $L_{10}^r(\mu_0)$ obtained by assuming the 
validity of the NLO representation of $\Delta\overline{\Pi}_{V-A}(Q^2)$
and using it to solve for $L_{10}^r$ at each $Q^2$. Results are shown 
for each of the four lightest $m_\pi$ ensembles ($E1-E4$). The measured 
values for the pseudoscalar masses and decay constants for the given 
ensemble~\cite{ainv231,ainv137} (see Table~\ref{tab:param}) are taken as 
inputs in all cases. Also shown, for comparison, are the results obtained 
from a similar NLO analysis of the continuum results. The uncertainties
on the continuum results (not shown explicitly) are small 
($\sim 2.5\%$) and strongly correlated in the region of $Q^2$ shown
in the figure.

While the incompatibility of the NLO form and the continuum results is 
immediately evident in the obvious non-constancy, within errors, of 
$L_{10}^r$ with respect to $Q^2$, it is far from clear that this would 
be the case if one had access only to the lattice results. In fact, if 
one imposes as input the (albeit non-conservative) assessment/assumptions 
of Ref.~\cite{gapp08} regarding ${\cal C}_0^r$ and ${\cal C}_1^r$, a NNLO 
fit does become possible, and returns a value for the NNLO LEC $C_{87}^r$ 
(which accounts for the bulk of the $Q^2$-dependence of 
$\Delta\overline{\Pi}_{V-A}(Q^2)$ in the low-$Q^2$ region) 
which is $\sim 2\sigma$ away from zero~\cite{boylelatt13vma}, showing
that the lattice data is capable of distinguishing, to some extent, 
between the NLO and NNLO forms. The lattice errors are, however,
much too large to allow a simultaneous fit of all four unknown LEC 
combinations $L_{10}^r$, $C_{87}^r$, ${\cal C}_0^r$ and ${\cal C}_1^r$. 

To make progress, a way must be found to combine
the lattice and continuum results, and take advantage of their complementary
strengths. We discuss a practical way of accomplishing this goal in
the next section.

\section{\label{sec5}Combining Lattice and Continuum Data to Improve
the Determination of $L_{10}^r$}
It is convenient to reduce the number of unknown LECs to be dealt
with by working with the difference of the physical-mass, continuum
and corresponding lattice results for $\Delta\overline{\Pi}_{V-A}(Q^2)$, 
evaluated at the same $Q^2$. With $L_9^r$ considered known~\cite{bt02}, 
the resulting difference
\begin{equation}
\delta \Delta\overline{\Pi}(Q^2)\equiv \left[\Delta\overline{\Pi}_{V-A}
(Q^2)\right]_{latt}\, -\, \left[\Delta\overline{\Pi}_{V-A}(Q^2)
\right]_{cont}\ ,
\label{defndeltadeltapibar}\end{equation}
depends only on the LECs $L_{10}^r$, ${\cal C}_0^r$ and ${\cal C}_0^r$.
Explicitly
\begin{equation}
\delta\Delta\overline{\Pi}(Q^2)\, =\, \Delta \hat{\cal R}^E(Q^2)
\, +\, \Delta c_{10}^E\, L_{10}^r\, +\, \delta^E_0 {\cal C}_0^r\, 
+\, \delta^E_1 {\cal C}_1^r\ ,
\label{deltadeltapibarnnlo}\end{equation}
where
\begin{eqnarray}
&&\Delta \hat{\cal R}^E(Q^2)\equiv 
\left[\hat{\cal R}_{NNLO}(Q^2)\right]_{latt}^E
\, -\, \left[\hat{\cal R}_{NNLO}(Q^2)\right]_{phys}\nonumber\\
&&\Delta c_{10}\equiv \left[ c_{10}\right]_{latt}^E-\left[ c_{10}\right]_{phys}
\nonumber\\
&&\delta_0\equiv \left[ m_\pi^2\right]_{latt}^E/\left[ m_\pi^2\right]_{phys}
\nonumber\\
&&\delta_1\equiv \left[ m_\pi^2+2m_K^2\right]_{latt}^E/
\left[ m_\pi^2+2m_K^2\right]_{phys}\ ,
\label{deltaRhatdefn}\end{eqnarray}
with the superscript $E$ labelling the ensemble under consideration
and the subscripts $phys$ and $latt$ indicating the values of the quantities
in question obtained using physical and lattice values for the relevant
pseudoscalar masses and decay constants, respectively. 
$\delta\hat{\cal R}^E(Q^2)$ and $\Delta c_{10}^E$
of course also depend on the chiral scale $\mu$.

\begin{figure}[H]
  \begin{minipage}[t]{0.44\linewidth}
{\rotatebox{270}{\mbox{
\includegraphics[width=0.9\textwidth]
{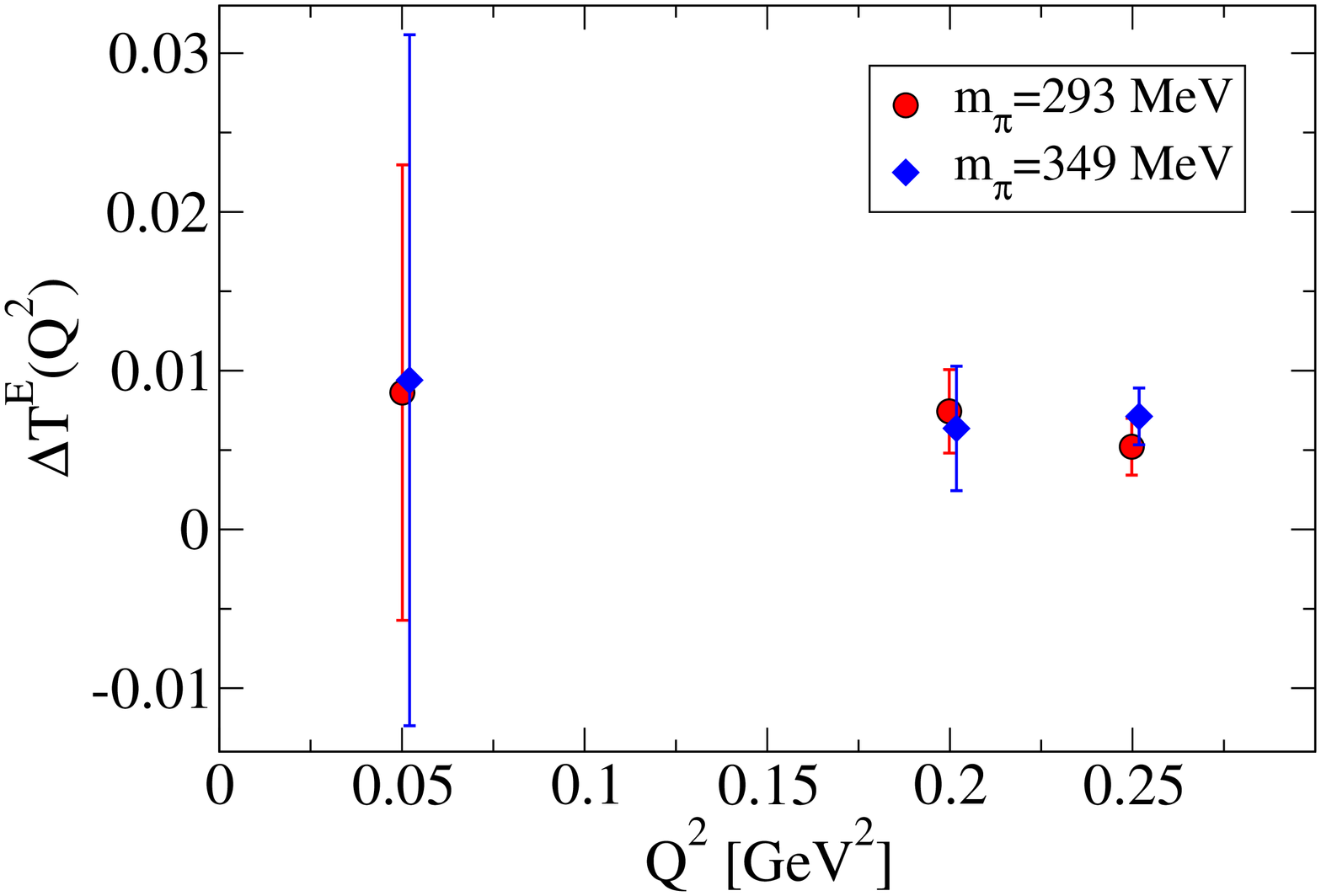}
}}}
  \end{minipage}
\hfill
  \begin{minipage}[t]{0.44\linewidth}
{\rotatebox{270}{\mbox{
\includegraphics[width=0.9\textwidth]
{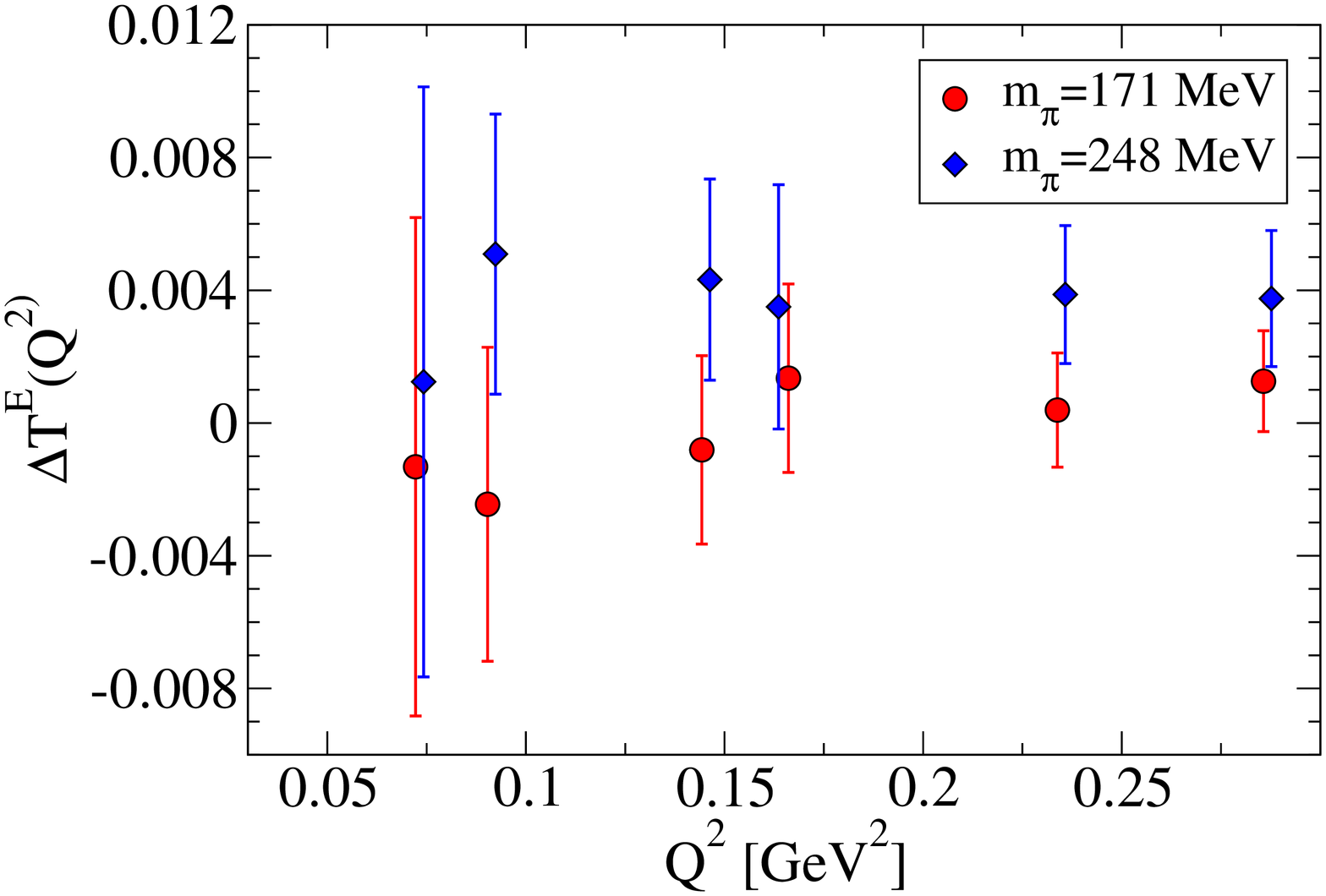}
}}}
  \end{minipage}
  \caption{\label{tq2fig}$\Delta T^E(Q^2)$ as a function of $Q^2$
for the $1/a=2.31\ GeV$ ensembles $E3$ ($m_\pi =293\ MeV$) and $E4$ 
($m_\pi =349\ MeV$) (left panel) and the $1/a=1.37\ GeV$ ensembles
$E1$ ($m_\pi =171\ MeV$) and $E2$ ($m_\pi =248\ MeV$) (right panel)}
\end{figure}

With this notation, the combined lattice-continuum constraints, 
for a given ensemble $E$, become
\begin{equation}
\Delta c_{10}^EL_{10}^r\, +\, \delta_0^E {\cal C}_0^r\, 
+\, \delta_1^E {\cal C}_1^r 
\, =\, \delta\Delta\overline{\Pi}(Q^2)\, -\, \Delta\hat{\cal R}^E(Q^2)
\, \equiv\, \Delta T^E(Q^2)\ .
\label{lattcontconstraintform}\end{equation}
Since both terms on the RHS are $Q^2$-dependent, while the LHS is
$Q^2$-independent, the versions of these constraints corresponding to
different $Q^2$, but the same lattice ensemble $E$ can be used to provide
checks on the self-consistency of the data employed, as well as on the 
reliability of the analysis framework. It turns out that the two constraints 
with reasonable errors obtained for the ensemble $E5$ do not pass this 
self-consistency test, while all of the available constraints are consistent 
for the other four ensembles. We thus exclude the ensemble $E5$ from the 
rest of the analysis. $E5$ is the ensemble with the largest pion mass, 
$m_\pi=399\ MeV$, a value which may, in any case, have been pushing
the bounds of the chiral analysis. The consistency of the constraints 
for the other four ensembles is displayed in Fig.~\ref{tq2fig}, which 
plots the $\Delta T^E(Q^2)$ 
for these ensembles for the $Q^2$ of interest to the chiral analysis. 
The left panel shows the results for the fine $1/a=2.31\ GeV$ 
ensembles $E3$ and $E4$, the right panel the results for the coarse 
$1/a=1.37\ GeV$ensembles $E1$ and $E2$. The lowest $Q^2$ points, 
at $Q^2=0.018\ GeV^2$, have been omitted from the right panel since
incorporating their absolutely enormous errors would force a dramatic 
increase in the range displayed on the vertical axis. In both panels, 
the $Q^2$ values for the ensemble with heavier value of $m_\pi$ have 
been shifted slightly to the right for presentational clarity.

For the remaining four ensembles employed in the analysis, a final combined 
version, $\Delta\overline{T}^E$, of the RHS of the constraint for 
each ensemble is obtained by performing a weighted average, over the 
points with $Q^2<0.3\ GeV^2$ available for that ensemble, of 
the corresponding $Q^2$-dependent RHSs. The average is more heavily 
weighted to the upper portion of the $Q^2$ analysis window, where the 
main source of error, that on $\Delta\overline{\Pi}_{V-A}(Q^2)$, is 
smaller, and, given the good self-consistency, we assign to 
$\overline{T}^E$ an uncertainty typical of the errors in this region.
The results of this exercise are
\begin{eqnarray}
&&\Delta\overline{T}^1\, =\, 0.0007(17)\nonumber\\
&&\Delta\overline{T}^2\, =\, 0.0039(21)\nonumber\\
&&\Delta\overline{T}^3\, =\, 0.0062(18)\nonumber\\
&&\Delta\overline{T}^4\, =\, 0.0070(18)\ .
\label{ensembleconstraints}\end{eqnarray}

Performing a combined fit incorporating the continuum 
$\Delta\overline{\Pi}_{V-A}(0)$ constraint, Eq.~(\ref{nnlopibarvmaq2eq0}), 
and the four lattice-continuum constraints obtained by employing the results
of Eqs.~(\ref{ensembleconstraints}) on the RHS of 
Eq.~(\ref{lattcontconstraintform}), we find
\begin{eqnarray}
&&L_{10}^r(\mu_0)\, =\, -0.0031(8)\nonumber\\
&&{\cal C}_0^r(\mu_0)\, =\, -0.0008(8)\nonumber\\
&&{\cal C}_1^r(\mu_0)\, =\, \ \ 0.014(11)\ .
\label{stage1results}\end{eqnarray}
The size of the
errors reflects the non-trivial size of the uncertainties
on the $\Delta\overline{T}^E$ in (\ref{ensembleconstraints}), and the fact 
that the associated constraints, (\ref{lattcontconstraintform}), are being 
required to provide information on two additional fit parameters. 
While the resulting errors, especially those on ${\cal C}_0^r$ and 
${\cal C}_1^r$, are larger than one might hope, they have,
at least, the advantage of being data-based.

The errors on the $\Delta\overline{T}^E$ in (\ref{ensembleconstraints}) 
result largely from those 
on the lattice data for $\Delta\overline{\Pi}_{V-A}(Q^2)$. It is,
unfortunately, difficult to significantly improve these, and
thus necessary to look to additional continuum input for any further
improvement. The existence of strong correlations amongst the fit 
parameters in (\ref{stage1results}) suggests that a single additional 
constraint should be sufficient to achieve a reduction
in the errors for all three fit parameters. Fortunately, such an 
additional constraint exists.

The source of this constraint is a recent IMFESR 
analysis~\cite{kmimfesr13} of the flavor-breaking (FB) correlator difference
\begin{equation}
\delta^{FB}\Delta\overline{\Pi}_{V-A}(Q^2)\, \equiv\,
\overline{\Pi}^{(0+1)}_{ud;V-A}(Q^2)-\overline{\Pi}^{(0+1)}_{us;V-A}(Q^2)\ ,
\label{fbimfestcorrelatordefn}\end{equation}
from which the result
\begin{equation}
\delta^{FB}\Delta\overline{\Pi}_{V-A}(0)\, =\, 0.0113(15)
\label{dkimfesrvmaudmusQ2eq0}\end{equation}
was obtained. The analysis employed (i) OPAL non-strange spectral data for 
the $V$ and $A$ channels~\cite{opalud99}, updated as in Ref.~\cite{dv72}; 
(ii) $us$ spectral data from ALEPH~\cite{alephus99} and the recent 
B-factory results for the exclusive mode $K^-\pi^0$~\cite{babarkmpi0}, 
$K_s\pi^-$~\cite{bellekspi}, $K^-\pi^+\pi^-$~\cite{babarkpipiallchg} and 
$K_s\pi^-\pi^0$~\cite{bellekspipi} invariant mass distributions measured in 
strange hadronic $\tau$ decays; and (iii) PDG~\cite{pdg12},
FLAG~\cite{flag}, and additional lattice~\cite{davies12,boyled2convlatt}
results for the treatment of, and input to, OPE contributions.
The $us$ exclusive mode distributions are normalized to current
strange $\tau$ branching fractions. 
We refer the reader to Ref.~\cite{kmimfesr13} for details of the analysis. 

The result given in Eq.~(\ref{dkimfesrvmaudmusQ2eq0}) is of interest 
for our purposes because the NNLO LEC contributions to the NNLO 
representation of $\delta^{FB}\Delta\overline{\Pi}_{V-A}(0)$ 
appear in precisely the combination ${\cal C}_0^r$. Explicitly,
\begin{equation}
\left[\delta^{FB}\Delta\overline{\Pi}_{V-A}(0)\right]_{NNLO}\, =\, 
{\cal R}^{FB}(0) + d^{FB}_5L_5^r + d^{FB}_9L_9^r
 + d^{FB}_{10}L_{10}^r + \left({\frac{m_K^2-m_\pi^2}{m_\pi^2}}
\right)\, {\cal C}_0^r\ ,
\label{nnlofbudmusvma}\end{equation}
where ${\cal R}^{FB}(Q^2)$ represents the sum of all 1- and 2-loop 
contributions with only LO vertices.
The (rather lengthy) expression for ${\cal R}^{FB}(0)$, as well as 
those for the $Q^2$-independent coefficients $d^{FB}_{5,9,10}$, are 
obtainable from the results quoted in Ref.~\cite{abt00} and not 
presented here. They are fully fixed once the chiral scale $\mu$ 
and pseudoscalar masses and decay constants are specified. 

Unlike the case of the NNLO representation of 
$\Delta\overline{\Pi}_{V-A}(0)$, where the coefficient $c_{10}$ of 
$L_{10}^r$ in Eq.~(\ref{udvmannlo}) contains both NLO and NNLO 
contributions, NLO contributions proportional to $L_{10}^r$ cancel in 
forming the FB difference $\delta^{FB}\Delta\overline{\Pi}_{V-A}(Q^2)$. 
The result is that $d^{FB}_{10}$ is purely NNLO, and suppressed 
numerically compared to $c_{10}$. The coefficient of ${\cal C}_0^r$
in Eq.~(\ref{nnlofbudmusvma}) is, in contrast, enhanced by 
the factor $(m_K^2-m_\pi^2)/m_\pi^2 \simeq 11.6$. The linear
combination of $L_{10}^r$ and ${\cal C}_0^r$ appearing in
(\ref{nnlofbudmusvma}) is thus very different from that appearing
in the continuum $\Delta\overline{\Pi}_{V-A}(0)$ constraint.
Since $L_9^r$ is well known~\cite{bt02}, and $L_5^r$, which is
also known~\cite{bj11}, is such that its contribution to the
RHS of (\ref{nnlofbudmusvma}) is numerically small, the 
result obtained by combining Eqs.(\ref{dkimfesrvmaudmusQ2eq0}),
and (\ref{nnlofbudmusvma}),
\begin{equation}
2.12\, L_{10}^r(\mu_0)\, +\, 11.6\, {\cal C}_0^r(\mu_0)\,
=\, -0.00346\, (161)\ ,
\label{udmusvmaconstraint}\end{equation}
provides the additional independent constraint we need.

We now have the two continuum constraints, Eqs.~(\ref{nnlopibarvmaq2eq0}) 
and (\ref{udmusvmaconstraint}), and four combined lattice-continuum 
constraints, Eq.~(\ref{lattcontconstraintform}). All of these can be cast 
in the form
\begin{equation}
a_{10}^{(k)}L_{10}^r\, +\, a_0^{(k)} {\cal C}_0^r\,
+\, a_1^{(k)} {\cal C}_1^r
\, =\, d^{(k)}\pm \delta d^{(k)}\, ,
\label{genericconstraintform}\end{equation}
with $k$ labelling the different constraints, the $a_{10}^{(k)}$,
$a_{0}^{(k)}$ and $a_{1}^{(k)}$ all known, and $\delta d^{(k)}$
the relevant error. For the four lattice-continuum constraints,
$\delta d^{(k)}$ is totally dominated by the error on the
lattice determination of the $\Delta\overline{\Pi}_{V-A}(Q^2)$
for the ensemble in question. For the continuum $V-A$
constraint, Eq.~(\ref{nnlopibarvmaq2eq0}), $\delta d^{(k)}$ is
determined by the experimental errors on the $ud$ $V-A$ spectral
distribution. Finally, for the FB continuum constraint,
Eq.~(\ref{udmusvmaconstraint}), $\delta d^{(k)}$ is 
dominated by the experimental errors on the $us$ spectral
distribution and $us$ $V/A$ separation uncertainties.
Since the dominant sources of error for the different constraints are 
independent, we fit $L_{10}^r(\mu_0)$, ${\cal C}_0(\mu_0)$ and
${\cal C}_1(\mu_0)$ by minimizing
\begin{equation}
\chi^2\, =\, \sum_k{\frac{\left[ d_k\, -\, \left( a_{10}^{(k)}\,
L_{10}^r(\mu_0)\, +\, a_0^{(k)}\, {\cal C}_0(\mu_0)\, +\, a_1^{(k)}\,
{\cal C}_1(\mu_0)\right)\right]^2}{\left[ \delta d^{(k)}\right]^2}}\ .
\label{chisqdefn}\end{equation}

Implementing this six-constraint fit, we find the significantly improved 
results
\begin{eqnarray}
&&L_{10}^r(\mu_0)\, =\, -0.00346\, (29)_{fit}(13)_{L_{5,9}^r}\nonumber\\
&&{\cal C}_0^r(\mu_0)\, =\, -0.00034\, (13)_{fit}(3)_{L_{5,9}^r}\nonumber\\
&&{\cal C}_1^r(\mu_0)\, =\, \ 0.0081(35)_{fit}(7)_{L_{5,9}^r}\ ,
\label{stage2results}\end{eqnarray}
where we have separated out the contributions to the errors from the 
uncertainties on the input values for $L_5^r(\mu_0)$ and $L_9^r(\mu_0)$. 
The resulting ${\cal C}^r_0(\mu_0)$-${\cal C}^r_1(\mu_0)$, 
${\cal C}^r_0(\mu_0)$-$L_{10}^r(\mu_0)$ and 
${\cal C}^r_1(\mu_0)$-$L_{10}^r(\mu_0)$ 
correlations are $-0.045$, $0.012$ and $0.978$, respectively.
The results (\ref{stage2results}) update the 
preliminary versions presented in Ref.~\cite{boyleetalvmalatt13},
and represent the best determination of $L_{10}^r$ to date{\footnote{The
reader might worry about the compatibility of the determination of 
$L_9^r(\mu_0)$ in Ref.~\cite{bt02}, our result for $L_{10}^r(\mu_0)$,
and the constraint on $L_9^r(\mu_0)+L_{10}^r(\mu_0)$ obtained from
the NNLO $SU(3)\times SU(3)$ analysis of radiative $\pi$ decay data, 
reported in Ref.~\cite{up08}. One should,
however, bear in mind that the latter constraint is obtained 
employing large-$N_c$ RChPT estimates for the NNLO LECs 
entering the axial amplitude from which the constraint is obtained.
In particular, central values of zero are used for all $1/N_c$-suppressed
LECs. It turns out that, as in the case of the continuum
$\Delta\overline{\Pi}_{V-A}(Q^2)$ constraint, a particular combination,
$4C_{13}^r+C_{64}^r+2\left( C_{13}^r-C_{62}^r+C_{81}^r\right)$, of 
$1/N_c$-suppressed NNLO LECs appears with a large ($2m_K^2/m_\pi^2\simeq 25$)
enhancement in its coefficient, relative to that of the 
non-$1/N_c$-suppressed NNLO LECs. We have, in fact, determined, as 
part of our fit, the $1/N_c$-suppressed combination 
$C_{13}^r(\mu_0)-C_{62}^r(\mu_0)+C_{81}^r(\mu_0)$. 
Shifting the central result $0$ used for this combination
in Ref.~\cite{up08} to the central value implied by our fit,
one finds a modified version of the radiative $\pi$ decay constraint on 
$L_9^r+L_{10}^r$ in excellent agreement with our result for
$L_{10}^r$ and that for $L_9^r$ in Ref.~\cite{bt02}. This exercise should, 
of course, be treated as illustrative only, since the discussion
makes no attempt to account for the effect of the additional, but unknown, 
$1/N_c$-suppressed combination $4C_{13}^r+C_{64}^r$. What it does
allow us to do, however, is conclude that the NNLO
$SU(3)\times SU(3)$ radiative $\pi$ constraint 
is subject to non-trivial uncertainties associated with contributions from
$1/N_c$-suppressed NNLO LECs, and, within these uncertainties,
perfectly compatible with our result for $L_{10}^r$.}}.

\section{\label{sec6}Summary and Discussion}
Our main results are those given in Eq.~(\ref{stage2results}), where the 
error labelled by the subscript $fit$ is that resulting from the errors 
on the two continuum and four lattice-continuum constraints employed in 
the combined fit. The key result is that for $L_{10}^r(\mu_0)$, though 
that for ${\cal C}_1^r(\mu_0)$ provides a further example of a NNLO LEC 
combination vanishing in the large-$N_c$ limit which cannot be neglected 
for $N_c=3$. 

It is worth commenting on the absence of constraints from the two
RBC/UKQCD ensembles with $1/a=1.75\ GeV$ in our analysis.{\footnote{For
further information on these ensembles, see Ref.~\cite{ainv175}.}}
These ensembles provide five $Q^2< 0.3\ GeV^2$, three with errors
small enough to be useful in assessing the self-consistency
of the $\Delta T^E(Q^2)$. The three low-error $\Delta T^E(Q^2)$ 
for the ensemble with $m_\pi =333\ MeV$, unfortunately, fail the 
self-consistency test.
Those for the ensemble with
$m_\pi =423\ MeV$ pass the self-cnsistency test, but correspond
to an $m_\pi$ which is both potentially rather large for use in an 
NNLO analysis and significantly larger than the largest value, 
$m_\pi =349\ MeV$, employed in the analysis discussed above. We 
can, however, use the results for the heavy $m_\pi$ ensemble to 
further test that the $m_\pi< 350\ MeV$ employed above lie
safely within the range of validity of the NNLO analysis framework.
To do so we have performed an extended version of the analysis
above, adding in the additional combined lattice-continuum 
constraint $\Delta \bar{T}^6=0.0048(17)$ obtained for the 
$1/a=1.75\ GeV$, $m_\pi =423\ MeV$ ensemble. The expanded fit 
yields results, ${\cal C}_0^r(\mu_0)\, =\, -0.0036(12)$,
${\cal C}_1^r(\mu_0)\, =\, 0.0070(24)$ and $L_{10}^r(\mu_0)\, =\, 
-0.00355(23)$, in excellent agreement with those of the 
main analysis. Since $m_\pi =423\ MeV$ is rather large,
we do not use the results of this extended analysis as
our main ones, but do argue that the stability of the results
with respect to such a large increase in the maximum $m_\pi$
employed provides strong evidence in support of the reliability
of our NNLO treatment of the lower-$m_\pi$ data. 

The only other NNLO determination of $L_{10}^r(\mu_0)$ we are aware of 
is that of Ref.~\cite{gapp08}. The central value in this case,
$L_{10}^r(\mu_0)\, =\, -0.00406(39)$, differs from ours by 
$\sim 2\sigma${\footnote{In terms of the error quoted
in Ref.~\cite{gapp08}, the difference in central values is
only $1.5\sigma$. Were the assumption used to generate it to be 
updated using the improved determination of ${\cal C}_0^r(\mu_0)$
obtained above, however, the error of Ref.~\cite{gapp08} would be 
reduced to $0.00023$. The determination of ${\cal C}_1^r$
using lattice data is key to bringing this type of difficult-to-quantify
uncertainty under control.}}. The difference results, essentially entirely, 
from the difference in ${\cal C}_1^r(\mu_0)$ values, with ${\cal C}_1^r(\mu_0)$
(then unknown) having been assigned the (assumed) central value $0$ 
in \cite{gapp08}, but fit, using lattice data, in our 
analysis{\footnote{A significant difference also exists between our 
result for ${\cal C}_0^r(\mu_0)$ and that used in Ref.~\cite{gapp08}. 
This results largely from an overestimate, by a factor of more than 
$2$~\cite{kmimfesr13}, in the RChPT value for $C_{80}^r(\mu_0)$ employed
in \cite{gapp08}. The smallness of the ${\cal C}_0^r$ contributions to 
the $\Delta\overline{\Pi}_{V-A}(0)$ constraint, however, means that 
this difference has a negligible impact on the results for 
$L_{10}(\mu_0)$.}}. The error on $L_{10}^r(\mu_0)$ in Ref.~\cite{gapp08}, 
as stressed in that reference, is completely dominated by the 
assumed uncertainty on ${\cal C}_1^r(\mu_0)$. This uncertainty is based
on the assumption that 
\begin{equation}
\vert C_{13}^r(\mu_0)-
C_{62}^r(\mu_0)+C_{81}^r(\mu_0)\vert\, <\, \vert C_{12}^r(\mu_0)-
C_{61}^r(\mu_0)+C_{80}^r(\mu_0)\vert /3\ ,
\label{gapp08assumption}\end{equation}
which turns out to be insufficiently conservative, and would
be even more so were the data-based result obtained above for 
${\cal C}_0^r(\mu_0)$ (which is $\sim -0.6$ times that 
employed in Ref.~\cite{gapp08}) to be used on the RHS. 
Our error has the advantage not only of being smaller, but of being
based entirely on lattice and continuum data errors and independent
of any additional assumptions.

It is useful to clarify the relative roles of the lattice-continuum 
and continuum constraint errors, since this determines where best to 
focus future efforts to further reduce the error on $L_{10}^r$. 
In this context, it is also relevant to bear in mind that the 
$\delta^{FB}\Delta\overline{\Pi}_{V-A}(0)$ constraint, 
Eq.~(\ref{udmusvmaconstraint}), which is crucial in achieving the 
reduced errors in (\ref{stage2results}), relies on current strange 
hadronic $\tau$ decay mode branching fractions for the normalizations 
of the exclusive strange mode contributions to the $us$ $V-A$ spectral 
function. These branching fractions, as well as the exclusive strange 
distributions, remain the subjects of ongoing experimental investigation. 
In addition, the $V/A$ separation of the exclusive $K\pi\pi$ mode spectral 
contributions, which is currently done only approximately, can, in 
principle, be improved through angular analyses~\cite{km92} which 
are feasible with B-factory data. Improvements to the FB IMFESR analysis, 
and hence to the associated FB $V-A$ constraint, are thus likely 
to be accessible in the near future.

In order to illustrate the impact plausible changes in
the $us$ $V-A$ spectral data might have on $L_{10}^r$, we have rerun 
the analysis described in Sec.~\ref{sec5} using as input to the FB $V-A$
IMFESR constraint, the alternate value, 
$\delta^{FB}\Delta\overline{\Pi}_{V-A}(0)\, =\, 0.0098(15)$, obtained in 
Ref.~\cite{kmimfesr13} using the alternate, still-preliminary BaBar results 
for the branching fractions $B[\tau^- \rightarrow K^-\, n\, \pi^0\nu_\tau ]$,
$n=1,2,3$, reported in Ref.~\cite{adametz}. The results of this exercise
are $L_{10}^r(\mu_0)\, =\, -0.00356(32)$, ${\cal C}_0^r(\mu_0)\, =\,
-0.00024(12)$ and ${\cal C}_1^r(\mu_0)\, =\, 0.0068(32)$. While the input
constraint value has been shifted by $1\, \sigma$, $L_{10}(\mu_0)$ has
shifted by only $\sim 1/3$ of the $fit$ component of the error in the 
main result. We learn from this exercise that, at present, it is the 
lattice errors on 
$\Delta\overline{\Pi}_{V-A}(Q^2)$ which dominate the uncertainty on 
$L_{10}^r$. Improvements in the error on the FB $V-A$ IMFESR constraint 
(the less precise of the two continuum constraints), though almost
certainly feasible in the near future, will not help to significantly
reduce the error on $L_{10}^r$. Further non-trivial improvement requires 
instead a reduction in the errors on the lattice determinations of 
$\Delta\overline{\Pi}_{V-A}(Q^2)$. A natural target in this regard is 
a reduction in the errors on the $\pi$ pole subtraction through a 
reduction in the errors on $f_\pi$ for the two coarse $1/a=1.37\ GeV$ 
ensembles, where these errors on the $f_\pi^2$ factor entering this
subtraction are currently a factor of $\sim 2.3$ larger than those for 
the fine $1/a=2.31\ GeV$ ensembles.

Our determination of $L_{10}^r$ allows us to also fix the
corresponding $SU(2)\times SU(2)$ LEC, $\ell_5^r$, whose relation to
$L_{10}^r$ at NNLO has been worked out in Ref.~\cite{ghis07}.
With $F_0$ the $\pi$ decay constant in the $SU(3)$ chiral limit,
$\hat{m}_K$ the $K$ mass in the limit $m_{u,d}\rightarrow 0$,
$\ell_K\equiv \log\left({\frac{\hat{m}_K^2}{\mu_0^2}}\right)$, 
$\nu_K\equiv {\frac{1}{32\pi^2}}\, \left(\ell_K+1\right)$
and $X\equiv {\frac{\hat{m}_K^2}{16\pi^2F_0^2}}$,
this relation takes the form~\cite{ghis07}
\begin{eqnarray}
\ell_5^r(\mu_0)\, =&&\, \left(1-2X\ell_K\right)\, L_{10}(\mu_0)
+{\frac{1}{12}}\nu_K\, +\, X\, (0.000339+0.002243\ell_K
-0.000396\ell_K^2)\nonumber\\
&&\qquad -\, X\, \ell_K\, L_9^r(\mu_0)\, -\, 8\hat{m}_K^2\,
\left[ C_{13}^r(\mu_0)-C_{62}^r(\mu_0)+C_{81}^r(\mu_0)\right]\ ,
\label{su3tosu2ghis}\end{eqnarray}
where, in writing the second line, we have converted the dimensionless 
versions of the NNLO LECs used in Ref.~\cite{ghis07} to the dimensionful
versions of Ref.~\cite{bce99} used above. Note that the last term
in this relation is proportional to the combination ${\cal C}_1^r$
determined above. Estimating $\hat{m}_K$ using the LO relation
$\hat{m}_K^2=\bar{m}_K^2-{\frac{1}{2}}m_\pi^2$ (with $\bar{m}_K$ the
average of the charged and neutral $K$ masses), and taking
$F_0\simeq 80\ MeV$ from the $n_f=2+1$ lattice results favored by the 
FLAG assessment~\cite{flag}, we obtain
\begin{equation}
\ell_5^r(\mu_0)\, =\, 1.430\, L_{10}(\mu_0)\, -\, 0.00046
\, +\, 0.215\, L_9^r(\mu_0) \, -\, {\frac{\hat{m}_K^2}{4(2m_K^2+m_\pi^2)}}\,
{\cal C}_1^r(\mu_0)\ .
\label{numsu3tosu2}\end{equation}
With the input of Ref.~\cite{bt02} for $L_9^r(\mu_0)$, we obtain,
taking into account the $0.978$ correlation between the fitted
values of $L_{10}^r(\mu_0)$ and ${\cal C}_1^r(\mu_0)$, 
\begin{equation}
\ell_5^r(\mu_0)\, =\, -0.00507(10)\ .
\label{ell5numvalue}\end{equation}
The uncertainty on $L_9^r(\mu_0)$ plays no role to the number of
significant figures quoted for the error. The result (\ref{ell5numvalue})
corresponds to the value
\begin{equation}
\bar{\ell}_5\, =\, 13.0\, (2)
\label{ell5bar}\end{equation}
for the scale-invariant coupling $\bar{\ell}_5$ defined in Ref.~\cite{gl84}.
This is not only in excellent agreement with the results
$\bar{\ell}_6=16.0(5)(7)$ and $\bar{\ell}_6-\bar{\ell}_5=3.0(3)$
quoted in Ref.~\cite{bijnenscd12}, arising from the NNLO $SU(2)\times SU(2)$
analyses of the $\pi$ vector form factor~\cite{bct98} and 
$\pi^+\rightarrow e^+\nu_e\gamma$~\cite{bt97}, respectively, but,
when combined with $\bar{\ell}_6-\bar{\ell}_5=3.0(3)$, in fact yields 
the somewhat improved determination $\bar{\ell}_6=16.0(4)$ for
$\bar{\ell}_6$.

We close by comparing our results to RChPT estimates for the 
LECs/LEC combinations determined in our analysis, RChPT being
the framework most often used to make such estimates in the 
literature. Large-$N_c$-based RChPT estimates~\cite{L10rLONc} for 
$L_{10}^r$ are scale-independent,
and usually taken to correspond to $\mu\simeq \mu_0$. The resulting
$L_{10}^r(\mu_0)$ ($\simeq -0.0054$) is significantly more negative
than indicated by our determination. The lack of scale-dependence in 
the large-$N_c$ version of the RChPT LEC predictions can be repaired
by going beyond leading order in $1/N_c$. This has been done for the 
$V-A$ correlator in Ref.~\cite{prsc08}, where $1/N_c$ corrections were
shown to lower the RChPT prediction for $L_{10}^r(\mu_0)$~\cite{prsc08}. 
The resulting prediction, with the scale-dependence now fully under 
control, is $-0.0044(9)$, compatible within errors with our result 
above. Large-$N_c$ RChPT predictions for the NNLO LECs entering the 
combination ${\cal C}_0^r$~\cite{abt00,up08,bj11,ceekpp05,km06} lead 
to a result ${\cal C}_0^r(\mu_0)\simeq -0.0004$, in good agreement 
with the result above. This agreement, however, results from a 
fortuitous cancellation, with RChPT predictions for the individual 
$C_{12}^r,\, C_{61}^r$ and $C_{80}^r$ differing significantly from 
the coupled channel dispersive result of Ref.~\cite{jopc12} for 
$C_{12}^r(\mu_0)$, and the results for $C_{61}^r(\mu_0)$ and 
$C_{80}^r(\mu_0)$ obtained in Ref.~\cite{kmimfesr13} using FB IMFESRs 
in combination with the results of our fit above. The large-$N_c$ 
RChPT prediction for the $1/N_c$-suppressed LEC combination ${\cal C}_1^r$ 
is, of course, zero. To the best of our knowledge, $1/N_c$-corrections
have not yet been investigated for any of the NNLO LECs.

\begin{acknowledgments}
The lattice computations were done using the STFC's DiRAC facilities at
Swansea and Edinburgh. PAB, LDD and RJH are supported by an STFC 
Consolidated Grant, and by the EU under Grant Agreement PITN-GA-2009-238353 
(ITN STRONGnet). EK was supported by the Comunidad Aut\`onoma de Madrid under
the program HEPHACOS S2009/ESP-1473 and the European Union under Grant
Agreement PITN-GA-2009-238353 (ITN STRONGnet). KM acknowledges the
hopsitality of the CSSM, University of Adelaide, and the
support of the Natural Sciences and Engineering Research Council of Canada.
JMZ is supported by the Australian Research Council grant FT100100005.
\end{acknowledgments}


\vfill\eject
\end{document}